\documentclass[aps,prl,reprint,secnumroman]{revtex4-2}
\usepackage[american]{babel}
\usepackage{amsfonts}
\usepackage{amsmath}
\usepackage{amssymb}
\usepackage{epsfig}
\usepackage{latexsym}
\usepackage{dsfont}

\usepackage{fancyhdr}
\usepackage{graphicx}

\usepackage{hyperref}
\usepackage{etex}
\usepackage{float}
\usepackage{slashed}
\usepackage{xcolor}

\usepackage{mathtools}% http://ctan.org/pkg/mathtools
\newcommand{\fsl}[1]{\ensuremath{\mathrlap{\!\not{\phantom{#1}}}#1}}
\usepackage{dcolumn} 
\usepackage{bm}
\usepackage{splitidx}

\usepackage{slashed}
\fancypagestyle{plain}{
	\lhead{}
	\fancyhead[R]{\thepage}
	\fancyhead[L]{}
	
	\fancyfoot{}
}

\makeatletter
\renewcommand{\c@secnumdepth}{0}
\makeatother

\newcommand{\id}{\mathds{1}}

\begin{document}
%%%%%%%%%%%%%%%%%%%%%%%%%%%%%%%%%%%%%%%%%%%%%%%%%%%

\title{Conformal gauge theory of vector-spinors and spin-$\tfrac{3}{2}$ particles}

\author{Dario Sauro}
\email{dario.sauro@uni-jena.de}
\affiliation{
Theoretisch-Physikalisches Institut, Friedrich-Schiller-Universität Jena,\\
Fröbelstieg 1, 07743 Jena, Germany
}

%%%%%%%%%%%%%%%
\begin{abstract}
%%%%%%%%%%%%%%%
%
The unique off-shell fermionic gauge invariance of a vector-spinor field theory is found, and the invariant action is derived. The latter is Weyl invariant in any dimension in the massless limit, and it coincides with the singular point of the one-parameter family of Rarita-Schwinger Lagrangians, in agreement with previous findings in flat space. Pure gauge configurations are represented by gamma-trace vector-spinors, which can be gauged away in a global fashion. Previous claims that this theory is classically inconsistent are shown to be flawed, and the Velo-Zwanziger instability is proved to be absent. The theory propagates a massive spin-$\tfrac{3}{2}$ particle together with a spin-$\tfrac{1}{2}$ state whose mass is twice that of the $j=\tfrac{3}{2}$ mode. The causal construction of the quantum field is consistent with the field equations in that the ratio of the masses is the same, while it shows that the lower-spin component is a negative-norm state. The conformal anomaly is derived using known results for the heat kernel of nonminimal second-order operators, and the resulting $a$ charge is negative consistently with the Hofman-Maldacena bound, which applies only to unitary theories. 
\end{abstract}

\maketitle

%\tableofcontents

\renewcommand{\thefootnote}{\arabic{footnote}}
\newcommand{\barpsi}{\overline{\psi}}
\newcommand{\barPsi}{\overline{\Psi}}
\newcommand{\barchi}{\overline{\chi}}
\newcommand{\barepsilon}{\overline{\epsilon}}
\newcommand{\barlambda}{\overline{\lambda}}
\setcounter{footnote}{0}

%%%%%%%%%%%%%%%%%%%%%%%%%%%%%%%%%%%%%%%%%%%%%%%%%%%%%%%%%%
\section{Introduction}\label{sect:intro}
%%%%%%%%%%%%%%%%%%%%%%%%%%%%%%%%%%%%%%%%%%%%%%%%%%%%%%%%%%

The most studied theory of spin-$\tfrac{3}{2}$ particles due to Rarita and Schwinger \cite{Rarita:1941mf} is known to lead to inconsistencies (the properties of such particles in supergravity are well studied yet out of the scope of this work). This field theory is described by a vector-spinor $\Psi_\mu$ whose vacuum field equations state the vanishing of the gamma-trace and longitudinal components, thus getting rid of the two spin-$\tfrac{1}{2}$ irreducible representations \cite{Das:1976ct}. As Johnson and Sudarshan \cite{Johnson:1960vt} first showed, a way to notice that the interacting theory is flawed is provided by computing the equal-time commutators and noticing that causality is jeopardized. Later, Velo and Zwanziger \cite{Velo:1969bt} proved that this feature reflects the appearance of a tachyonic mode, which is found by analyzing the normals to the characteristic surfaces of the equations of motion \emph{after} substitution of the primary and secondary constraints. Heuristically, this problem arises due to an accidental symmetry of the free theory, which is broken when coupled to external backgrounds such as the electromagnetic field. This implies that the number of propagating degrees of freedom (d.o.f.)\ is different in the free and interacting theories.

The action introduced in \cite{Rarita:1941mf} actually belongs to a one-parameter family of actions which are physically indistinguishable, and the parameter generates the so-called contact symmetry, see, e.g.,\ \cite{Pilling:2004cu}. In particular, the construction of \cite{Pilling:2004cu} shows that this theory can be derived by demanding that there are $8$ propagating degrees of freedom carried by a massive spin-$\tfrac{3}{2}$ particle. Notice that there is a singular value of this parameter in which the aforementioned analyses do not apply anymore. Furthermore, in a recent paper \cite{Valenzuela:2023aoa} the authors performed a Hamiltonian analysis of the constraints of the theory, showing that Dirac's conjecture \cite{Dirac:1950} that the secondary constraints ``generate" gauge transformations does not apply for the Rarita-Schwinger model, and also pointing out that there is a \emph{physical} spin-$\tfrac{1}{2}$ excitation.

In this paper the field theory of vector-spinors is reconsidered starting from a different perspective. The inception is the following question: are there any consistent (i.e.,\ not accidental) gauge symmetries that can be imposed at the level of the kinetic Lagrangian? The answer is that there is a unique such symmetry, which can also be imposed on the mass terms, thus yielding a gauge invariant massive theory. The action matches the singular value of the one-parameter family studied in \cite{Haberzettl:1998rw,Anselmi:1999bu,Anselmi:2020opi}, which is interpreted in a novel manner. By analyzing the conformal (Weyl) properties of vector-spinor theories it is noticed that the massless theory is not only conformal (like in the flat-space limit \cite{Anselmi:1999bu}), by it is the \emph{only} theory belonging to the intersection of conformal actions and the aforementioned one-parameter family. Velo and Zwanziger's computation \cite{Velo:1969bt} is carried over in this model too, yielding no superluminal propagation for any configuration of the external gauge potentials.

By using spin projectors one deduces that the gamma-trace vector-spinors belong to the kernel of the action functional and can be gauged away in a global fashion. This is the reason why it was mistakenly stated that the Green's function does not exist in the ``singular" limit of the one-parameter family of operators. The resulting field equations are different from those encountered in the literature. Specifically, in a local patch and in absence of interactions the equations of motion can be written as two Dirac equations for the transverse and longitudinal components of a gamma-traceless vector-spinor $\psi_\mu$. The physical prediction is that the spin-$\tfrac{1}{2}$ mode has twice the mass of the spin-$\tfrac{3}{2}$ particle.

The quantization problem of the physical gamma-traceless vector-spinor $\psi_\mu$ is tackled by following Weinberg's construction of causal quantum fields \cite{Weinberg:1964cn} (see also Chapter $5$ of \cite{Weinberg:1995mt}). The mode decomposition and the relative normalization of the $j=\tfrac{3}{2}$ and $j=\tfrac{1}{2}$ coefficient functions are derived by requiring that all anticommutators vanish at spacelike separations. Most importantly, the same field equations derived from first principles arise as a by-product of the construction, and the Feynman propagator is explicitly built. On the other hand, the lower-spin component is found to be a ghost, in agreement with previous analyses \cite{Anselmi:1999bu,Anselmi:2020opi}.

Finally, the $1$-loop quantum fluctuations of the field $\psi_\mu$ on top of a general Riemannian background are considered by computing the logarithmically divergent part of the effective action using the heat kernel technique. This is achieved by employing the recently found model-independent expression for the traced second heat kernel coefficient for nonminimal second-order operators \cite{Sauro:2025sbt}. The result is consistent with that of a Weyl conformal theory explicitly broken by the presence of a mass term in the action. In this case the physical prediction is that the contribution to the $a$-anomaly \cite{Hofman:2008ar,Komargodski:2011vj} has an opposite sign with respect to fields with spin $s\leq1$.

The Appendix fixes the conventions regarding the spacetime signature and the Clifford algebra, provides the explicit forms of the coefficient functions that are not reported in the main text, and contains the expressions of some tensor coefficients that appear in the heat kernel integration of the $1$-loop fluctuations.

\section{The Rarita-Schwinger framework}\label{sect:rarita-schwinger}

Here we briefly introduce the Rarita-Schwinger framework for describing spin-$\tfrac{3}{2}$ particles through a specific action functional involving the vector-spinor $\Psi_\mu$. To derive this action one usually requires it to be at most linear in the derivatives, Hermitian and such that the equations of motion of the free theory can be written as \cite{Das:1976ct,Fang:1979hq,Pilling:2004cu,Francia:2006hp}
\begin{equation}\label{eq:RS-field-equations}
	\begin{split}
	& (\fsl{\partial} + m) \Psi_\mu = 0 \, ,\\
	& \gamma \cdot \Psi = 0 \,  .
	\end{split}
\end{equation}
By combining the two equations one notices that
\begin{equation}
	\partial \cdot \Psi = 0 \, .
\end{equation}
It turns out that there exists a one-parameter family of Lagrangians that yields these field equations. The operator that enters such Lagrangians depends on a free real parameter $a$ which tunes an internal rotation in the spin-$\tfrac{1}{2}$ subspace, and it reads \cite{Pilling:2004cu}
\begin{equation}\label{eq:one-parameter-operator}
	\begin{split}
		\Lambda^{\mu\nu}(a) = & \left( \fsl{\nabla} + m  \right) g^{\mu\nu} + a \left( \gamma^\mu \nabla^\nu + \nabla^\mu \gamma^\nu \right)\\
		& + \frac{(d-1)a^2 + 2 a +1}{d-2} \gamma^\mu \fsl{\nabla} \gamma^\nu\\
		& - \frac{d(d-1)a^2+4(d-1)a+d}{(d-2)^2} m \gamma^\mu \gamma^\nu \, .
	\end{split}
\end{equation}
This internal rotation signals an underlying contact symmetry of the Lagrangian. The operator that is applied to a generic vector-spinor $\Psi_\mu$ and rotates the two spin-$\tfrac{1}{2}$ eigenspaces is
\begin{equation}\label{eq:rotation}
	\theta_\mu{}^\nu(b) = \delta_\mu{}^\nu + b \, \gamma_\mu \gamma^\nu \, ,
\end{equation}
which is singular for $b=-\tfrac{1}{d}$. Recalling the form of the Rarita-Schwinger operator \cite{Rarita:1941mf,Velo:1969bt,Das:1976ct}
\begin{equation}\label{eq:RS-operator}
	\Lambda^{\mu\rho}_{RS} = \gamma^{\mu\nu\rho} \nabla_\nu - m \, \gamma^{\mu\rho} \, ,
\end{equation}
we can write down the one-parameter operator Eq.\ \eqref{eq:one-parameter-operator} as
\begin{equation}
	\Lambda^{\mu\nu}(a) = \theta^\mu{}_\rho (-\tfrac{a+1}{d-2}) \, \Lambda^{\rho\lambda}_{RS} \, \theta_\lambda{}^\nu (-\tfrac{a+1}{d-2}) \, .
\end{equation}
The properties of the system are insensitive to the value of the free parameter $a$, which bears no physical meaning. The rotations in Eq.\ \eqref{eq:rotation} give rise to a group structure, and the symmetry of the physical properties under such rotations is known as contact symmetry \cite{Pilling:2004cu}. On the other hand, we notice that for $a=-\tfrac{2}{d}$ the parameter of the rotation assumes its singular value $b= - \tfrac{1}{d}$, and the physics described by the theory is completely different.

Whenever $a\neq-\tfrac{2}{d}$ the field equations $R^\mu=0$ have both a primary constraint $\gamma \cdot R=0$ and a secondary constraint $\nabla \cdot R=0$. In the free theory in flat spacetime this implies that both spin-$\tfrac{1}{2}$ modes of $\Psi_\mu$ are unphysical. However, in the presence of nontrivial background electromagnetic fields one can see that one spin-$\tfrac{1}{2}$ mode \emph{is} physical \cite{Velo:1969bt}. Thus, an inconsistency arises in that there is a difference in the number of the dynamical degrees of freedom between the free and interacting theories. Furthermore, in the latter case an unphysical superluminal propagation of the vector-spinor is found along a specific direction, thus providing a nontrivial example of how Lorentz invariance can be broken in spite of the manifest covariance of the field equations \cite{Velo:1969bt}. Technically speaking, this behavior arises because both constraints can be substituted in the field equations in a local fashion, so that superluminal propagation is seen to occur in the presence of a static homogeneous magnetic field by studying the normals to the characteristic surfaces. This unphysical property can be attributed to the fact that the longitudinal modes $\nabla_\mu \epsilon$ belong to the kernel of the Rarita-Schwinger operator Eq.\ \eqref{eq:RS-operator} in the free field limit only, while they contribute to the dynamics whenever general background fields are present.

\section{Fermionic gauge invariance}\label{sect:ferm-gauge-transf}

The most general linear self-adjoint differential operator acting on a vector-spinor $\Psi_\mu$ is given by
\begin{align}\label{eq:general-lin-diff-op}
 \Lambda^{\mu\rho} = \gamma^{\mu\nu\rho} \nabla_\nu + \alpha \left( \gamma^\mu \nabla^\rho + \gamma^\rho \nabla^\mu \right) + \beta \gamma^\mu \fsl{\nabla} \gamma^\rho \, .
\end{align}
We want to constrain the two dimensionless coefficients $\alpha$ and $\beta$ by imposing a fermionic gauge invariance like the one considered in \cite{Rarita:1941mf,Das:1976ct}, but requiring that this symmetry holds off the mass shell and in the presence of arbitrary interactions. A gauge transformation is written as a linear differential operator acting on a Dirac spinor $\epsilon$. Thus, the most general one comprises two independent tensor structures, i.e., 
\begin{equation}\label{eq:general-ferm-gauge-transf}
 \delta^{F}_{\theta_{1,2}} \Psi_\mu = \theta_1 \nabla_\mu \epsilon + \theta_2 \gamma_\mu{}^\nu \nabla_\nu \epsilon \, .
\end{equation}
Here $\theta_{1,2}$ are real constants, and the commonly studied case is given by $\theta_2=0$ \cite{Rarita:1941mf,Das:1976ct}. By imposing gauge invariance, i.e.,
\begin{equation}
 \delta^F_{\theta_{1,2}} \Lambda^{\mu\rho} \Psi_\rho \overset{!}{=} 0 \, ,
\end{equation}
we find that there is a single solution to the previous equation, which sets $\theta_1=\theta_2$, i.e.,
\begin{equation}\label{eq:gauge-transf}
\delta^F_\epsilon \Psi_\mu = \nabla_\mu \epsilon + \gamma_\mu{}^\nu \nabla_\nu \epsilon = \gamma_\mu \fsl{\nabla} \epsilon \, ,
\end{equation}
whereas the invariant operator is
\begin{align}\label{eq:gauge-inv-kin-term}\nonumber
 \Lambda^{\mu\rho}_{\rm inv} = & \, \gamma^{\mu\nu\rho} \nabla_\nu + \frac{d-2}{d} \left( \gamma^\mu \nabla^\rho + \gamma^\rho \nabla^\mu \right)\\\nonumber
 & - \frac{(d+1)(d-2)}{d^2} \gamma^\mu \fsl{\nabla} \gamma^\rho \, .
\end{align}
In contrast with the local transformation $\delta \psi_\mu = \nabla_\mu \epsilon$, Eq.\ \eqref{eq:gauge-transf} does not require any \emph{ad hoc} assumption about the background fields to yield an invariant action. Moreover, the local transformation \eqref{eq:gauge-transf} also permits to write down a gauge invariant mass term
\begin{equation}
 \Lambda^{\mu\rho}_m = \frac{m(d-1)}{d} \left( g^{\mu\rho} - \frac{1}{d-1} \gamma^{\mu\rho} \right) \, .
\end{equation}
Thus, using Eq.\ \eqref{eq:gamma3-to-gamma1} the massive gauge invariant operator can be written as
\begin{equation}\label{eq:gauge-inv-linear-differential-operator}
	\begin{split}
		\Lambda^{\mu\nu} = & \, \fsl{\nabla} g^{\mu\nu} - \frac{2}{d} \left( \gamma^\mu \nabla^\nu + \nabla^\mu \gamma^\nu \right) - \frac{d+2}{d^2} \gamma^\mu \fsl{\nabla} \gamma^\nu\\
		& + \frac{m(d-1)}{d} \left( g^{\mu\nu} - \frac{1}{d-1} \gamma^{\mu\nu} \right) \, .
	\end{split}
\end{equation}
This operator corresponds to the ``singular" case of the one-parameter family of operators, see Eq.\ \eqref{eq:one-parameter-operator}. However, fermionic gauge invariance makes this operator \eqref{eq:gauge-inv-linear-differential-operator} stand on a new and independent footing. As a matter of fact, the previously mentioned impossibility of deriving the Green's function \cite{Pilling:2004cu} is manifestly due to gauge invariance. Thus, if one is to find such a Green's function, a gauge-fixing term should be added to the action, whose operator may be written as
\begin{equation}\label{eq:gauge-fixing-linear-operator}
	\Lambda_{\rm gf}^{\mu\nu} = \frac{1}{\xi} \gamma^\mu \fsl{\nabla} \gamma^\nu \, .
\end{equation}

\subsection{Conformal properties}

Let us now analyze the conformal (Weyl) properties of our operator Eq.\ \eqref{eq:gauge-inv-linear-differential-operator}. First of all, we fix our conventions by choosing
\begin{eqnarray}
	\delta^W_\sigma g_{\mu\nu} = 2 \sigma g_{\mu\nu} \, , && \qquad  \delta^W_\sigma \Psi_\mu = w \sigma \Psi_\mu \, \, ,
\end{eqnarray}
where $w$ is the Weyl's weight of the vector-spinor, which is fixed by noticing that the infinitesimal transformation of curved-space gamma matrices is
\begin{equation}
	\delta^W_\sigma \gamma_\mu = \sigma \gamma_\mu  \, .
\end{equation}
Here, the weight is $w_{\Psi_\mu} = - \tfrac{d-3}{2}$. Moreover, the torsion-free spin connection $\omega^{ab}{}_\mu$ and the affine connection $\Gamma^\rho{}_{\nu\mu}$ transform as \cite{Sauro:2022chz}
\begin{equation}
	\begin{split}
		& \delta^W_\sigma \omega^{ab}{}_\mu = \left( e^a{}_\mu E^{\nu \,b} - e^b{}_\mu E^{\nu \,a} \right) \partial_\nu \sigma   \, ;\\
		& \delta^W_\sigma \Gamma^\rho{}_{\nu\mu} = 2 \delta^\rho{}_{(\nu} \partial_{\mu)} \sigma - g_{\mu\nu} \partial^\rho \sigma \, .
	\end{split}
\end{equation}
Up to homogeneous terms the Weyl transformations of the three different kinetic tensor structure operators that enter Eq.\ \eqref{eq:gauge-inv-linear-differential-operator} are
\begin{equation}
	\begin{split}
		& \delta^W_\sigma \gamma^{\mu\nu\rho} \nabla_\nu \Psi_\rho = (d-2) \left( \gamma^{[\rho} \partial^{\mu]} \sigma \right)  \Psi_\rho ;\\
		& \delta^W_\sigma \left( \gamma^\mu \nabla^\rho + \gamma^\rho \nabla^\mu \right) \Psi_\rho = - d \left( \gamma^{[\rho} \partial^{\mu]} \sigma \right) \Psi_\rho ;\\
		& \delta^W_\sigma \gamma^\rho \fsl{\nabla} \gamma^\mu \Psi_\mu = 0  \, ,
	\end{split}
\end{equation}
Therefore, the newly found fermionic gauge transformation \eqref{eq:gauge-transf} is compatible with Weyl invariance when $m=0$, whereas this is not the case for the commonly studied local transformation $\delta \Psi_\mu = \nabla_\mu \epsilon$. Furthermore, the kinetic part of Eq.\ \eqref{eq:gauge-inv-linear-differential-operator} represents the only element of the intersection between conformal operators and the one-parameter family of operators \eqref{eq:one-parameter-operator}. Notice also that Weyl invariance is not sufficient for fixing the relative coefficients of the three differential operators entering Eq.\ \eqref{eq:general-lin-diff-op}, for the last term in \eqref{eq:general-lin-diff-op} is Weyl covariant. Thus, the conformal operator for a vector-spinor $\Psi_\mu$ can be written in terms of a dimensionless constant $c$ as
\begin{equation}
	\Lambda_{\rm conf} = \fsl{\nabla} g^{\mu\nu} - \frac{2}{d} \left( \gamma^\mu \nabla^\nu + \nabla^\mu \gamma^\nu \right) + c \, \gamma^\mu \fsl{\nabla} \gamma^\nu  \, ,
\end{equation}
Thus, the compatibility of fermionic and Weyl gauge invariances manifestly holds in \emph{any} spacetime dimension, while an analogous feature arises in electrodynamics and Yang-Mills theories only in $d=4$. In spite of this appealing feature we shall not attempt to study the physics of conformal vector-spinors in a generic dimension $d$.

\section{Consistency of the classical theory}\label{sect:classical-consistency}

We now turn to analyzing the propagating degrees of freedom of the fermionic gauge invariant action \eqref{eq:gauge-inv-linear-differential-operator}. To this end, we first employ spin projectors to rewrite the operator Eq.\ \eqref{eq:gauge-inv-linear-differential-operator} in a more convenient form and derive the field equations, then we analyze the constraints that follow from the field equations and finally we show that no Velo-Zwanziger instability is present.

\subsection{Spin projectors}\label{subsect:spin projectors}

We choose a different parametrization of the spin projectors onto the spin-$\tfrac{1}{2}$ sector with respect to that which has been employed in the literature \cite{VanNieuwenhuizen:1981ae,Pilling:2004cu}. The first such projector is given by
\begin{equation}\label{eq:spin-proj1/2AA}
	\left(P^{\frac{1}{2}}_{AA}\right)_{\mu\nu} = \frac{1}{d} \gamma_\mu \gamma_\nu \, .
\end{equation}
which projects onto vector-spinors that can be written as $\gamma_\mu$ times a Dirac spinor. The complementary spin-$\frac{1}{2}$ projector is given by
\begin{align}\label{eq:spin-proj1/2BB}
	\left(P^{\frac{1}{2}}_{BB}\right)_{\mu\nu} = \frac{1}{p^2(d-1)} & \left[ d \, p_\mu p_\nu -  \left( \gamma_\mu \fsl{p} p_\nu + p_\mu \fsl{p} \gamma_\nu \right) \right.\\\nonumber
	& \left. + \frac{p^2}{d} \gamma_\mu \gamma_\nu \right] \, ,
\end{align}
which produces gamma-traceless longitudinal vector-spinors. The intertwining projectors are
\begin{equation}\label{eq:spin-proj1/2-intertwinings}
	\begin{split}
		& \left(P^{\frac{1}{2}}_{AB}\right)_{\mu\nu} = \frac{1}{p^2\sqrt{d-1}} \left[ \frac{p^2}{d} \gamma_\mu \gamma_\nu - \gamma_\mu \fsl{p}  p_\nu  \right] \, ;\\
		& \left(P^{\frac{1}{2}}_{BA}\right)_{\mu\nu} = \frac{1}{p^2\sqrt{d-1}} \left[ \frac{p^2}{d} \gamma_\mu \gamma_\nu - p_\mu  \fsl{p} \gamma_\nu  \right] \, .
	\end{split}
\end{equation}
Finally, the spin-$\tfrac{3}{2}$ projector reads
\begin{align}\label{eq:spin-proj3/2}
	\left(P^{\frac{3}{2}}\right)_{\mu\nu} = \frac{1}{p^2(d-1)} & \left[ (d-1) p^2 g_{\mu\nu} - d \, p_\mu p_\nu \right.\\\nonumber
	& \left. +  \left( \gamma_\mu \fsl{p} p_\nu + \gamma_\nu \fsl{p} p_\mu \right) - \gamma_\mu \gamma_\nu p^2 \right] \, ,
\end{align}
and the usual completeness and idempotence relations are satisfied
\begin{equation}
	\begin{split}
		& \left(P^{\frac{3}{2}} + P^{\frac{1}{2}}_{AA} + P^{\frac{1}{2}}_{BB}\right)_{\mu\nu} = g_{\mu\nu} \, ;\\
		& \sum_{i=A,B} P^{\frac{1}{2}}_{ji} P^{\frac{1}{2}}_{ik} = P^{\frac{1}{2}}_{jk} \, ;\\
		& P^{\frac{3}{2}} P^{\frac{3}{2}} = P^{\frac{3}{2}} \, .
	\end{split}
\end{equation}
Furthermore, it turns out convenient to define the projector onto gamma-traceless vector-spinors, which reads
\begin{equation}\label{eq:def-gamma-traceless-proj}
	\left(P^{\gamma T}\right)_{\mu\nu} \equiv \left(P^{\frac{3}{2}}\right)_{\mu\nu} + \left(P^{\frac{1}{2}}_{BB}\right)_{\mu\nu} = g_{\mu\nu} - \frac{1}{d} \gamma_\mu \gamma_\nu  \, .
\end{equation}
Let us point out that both projectors \eqref{eq:spin-proj1/2AA} and \eqref{eq:def-gamma-traceless-proj} are \emph{ultralocal}, i.e., they do not depend on the momentum $p^\mu$. 

\subsection{Field equations and constraints}\label{subsect:field-equations}

The first step in the analysis of the dynamical degrees of freedom is provided by rewriting the gauge invariant operator Eq.\ \eqref{eq:gauge-inv-linear-differential-operator} using the spin projector Eq.\ \eqref{eq:def-gamma-traceless-proj}. The result is remarkably simple, i.e,
\begin{align}
	\Lambda^{\mu\nu} = \left(P^{\gamma T}\right)^{\mu}{}_\rho \left( \fsl{\nabla} + m \right) \left(P^{\gamma T}\right)^{\rho\nu} \, .
\end{align}
Thus, gauge invariance tells us that the gamma-trace vector-spinor configurations belong to the kernel of the action. Indeed, by performing the following \emph{global} splitting
\begin{equation}\label{eq:gamma-trace-splitting}
	\Psi_\mu = \psi_\mu + \frac{1}{d} \gamma_\mu \chi \, , \qquad {\rm with} \quad \chi \equiv \gamma^\mu \Psi_\mu 
\end{equation}
we find that
\begin{equation}\label{eq:gauge-fixed-action}
	 S[\barPsi,\Psi] = S[\barpsi,\psi] = \int \barpsi{}^\mu \left( \fsl{\nabla} + m \right) \psi_\mu \, .
\end{equation}
Thus, the number of degrees of freedom in $d=4$ is at most $16-4=12$.

To find the field equations we have to remember that the gamma-traceless condition is purely algebraic, i.e., it is analogous to the traceless condition for symmetric $2$-tensors. Hence, functional variations must be consistent with such an algebraic property
\begin{equation}
	\frac{\delta \barpsi_\mu}{\delta \barpsi_\nu} = \delta^\nu{}_\mu - \frac{1}{d} \gamma^\nu \gamma_\mu \, .
\end{equation}
Accordingly, by varying Eq.\ \eqref{eq:gauge-fixed-action} we find the following field equations for the gamma-traceless vector-spinor $\psi_\mu$
\begin{align}\label{eq:field-equations}
	& R_\mu \equiv \fsl{\nabla} \psi_\mu - \frac{2}{d} \gamma_\mu \nabla \cdot \psi + m \, \psi_\mu = 0 \, .
\end{align}
In the Rarita-Schwinger framework the gamma-traceless condition is derived as a constraint following from the vacuum field equations by computing $\gamma \cdot R = 0$, whereas now $\gamma \cdot R$ vanishes identically. Thus, only the secondary constraint is nontrivial, and it reads
\begin{align}\label{eq:secondary-constraint}
	\nabla \cdot R = & \, \frac{d-2}{d} \left( \fsl{\nabla} + \frac{d}{d-2} m \right) \nabla \cdot \psi \\\nonumber
	& + \gamma^\lambda \left( \frac{1}{2} R^\mu{}_\lambda + i e F^\mu{}_\lambda \right) \psi_\mu = 0 \, ,
\end{align}
where $e$ is the charge of the vector-spinor. Another striking difference with respect to the Rarita-Schwinger theory arises from by analyzing the secondary constraint. In flat space and in the absence of background electromagnetic fields, Eq.\ \eqref{eq:secondary-constraint} tells us that the longitudinal mode $\nabla \cdot \psi$ obeys the Dirac equation with mass $\frac{d}{d-2} m$, while in the Rarita-Schwinger $\nabla \cdot \psi =0$ in such a limit, yielding $8$ degrees of freedom.

\subsection{Absence of Velo-Zwanziger instabilities}

To proceed further, let us specialize to the physical dimension $d=4$ and briefly perform a $3+1$ decomposition of the Lagrangian. Gauge invariance then implies that only $12$ d.o.f.\ are physical, and we can exploit the gamma-tracelessness to write
\begin{equation}
	\psi_0 = \gamma^0 \gamma^i \psi_i \, .
\end{equation}
Accordingly, the Lagrangian ${\cal L}$ takes the following form
\begin{equation}\label{eq:3+1-lagrangian}
	{\cal L} = - \barpsi_j \left[ \delta^{ij} \left( \fsl{\nabla} + m \right) + \gamma^j \left( \fsl{\nabla} + m \right) \gamma^i  \right] \psi_i \, .
\end{equation}
In contrast with the analysis performed by Velo and Zwanziger \cite{Velo:1969bt} we cannot make use of the secondary constraint Eq.\ \eqref{eq:secondary-constraint} in a local fashion. Thus, the previous Lagrangian already yields the first-order differential operator whose normals $n_\mu$ to the characteristic surfaces we want to study. This is done by extracting the symbol $D^{ji}$ that multiplies the time derivative in the previous equation, i.e.,
\begin{equation}\label{eq:characteristic-surface-matrix}
	D^{ji} = \delta^{ji} \gamma^0 + \gamma^j \gamma^0 \gamma^i \, .
\end{equation}
Notice that the $8$-dimensional space defined by $\gamma^i\psi_i=0$ provides eigenvectors of the previous matrix with eigenvalue $\gamma^0$, whereas the $4$-dimensional space spanned by $\psi_i = \gamma_i \lambda$ also yields eigenvectors, whose eigenvalues are $-2\gamma^0$. The physical solutions of the field equations are also solutions of
\begin{equation}
	\det D^{ij} = 0 \, ,
\end{equation}
which determines the normals to the characteristic surfaces. Using a general normal vector $n_\mu$ and employing the previous argument we can rewrite the last equation in a Lorentz covariant fashion as
\begin{equation}
	\det D = \left( n_\mu \gamma^\mu \right)^8  \, \left( 2 n_\mu \gamma^\mu \right)^4 = 2^2 \left(n^2 \right)^6 = 0 \, .
\end{equation}
Clearly, there exists no configuration of the external gauge fields such that this equation is solved for spacelike $n_\mu$. Therefore, the loss of causality that was found in the Rarita-Schwinger case \cite{Velo:1969bt} does not take place here. Therefore, in $d=4$ the field equations in Eq.\ \eqref{eq:field-equations} describe the causal propagation of $8$ spin-$\tfrac{3}{2}$ degrees of freedom with mass $m$ and $4$ spin-$\tfrac{1}{2}$ d.o.f.\ with mass $2m$.

\section{Quantum theory: the mode decomposition}\label{sect:mode-decomposition}

A free massive quantum field theory is defined by the eigenvalues of the Casimir of the Poincarè algebra, which are described by the mass $m$ and spin $j$. These identify particles together with additional quantum numbers, such as the electric charge. For massive theories the intrinsic redundancy of the field theoretical description is removed by noting that particles must provide irreducible representations of Wigner's little group, i.e., spatial rotations \cite{Weinberg:1964cn}. Furthermore, fields with a nonzero number of spacetime indices may carry more than a single physical such irreducible representation, thus describing more than a single physical particle (see, e.g., \cite{Percacci:2025oxw}).

Let us illustrate this feature of quantum field theory in the case of a gamma-traceless vector-spinor
\begin{equation}\label{eq:vector-spinor-irrep}
	\psi_\mu \in \left(1, \tfrac{1}{2} \right) \oplus \left(\tfrac{1}{2},1 \right) \, , \qquad \gamma^\mu \psi_\mu = 0 \, ,
\end{equation}
which belongs to a $12$-dimensional irreducible representation of the Lorentz group. A complex spin $j=\tfrac{3}{2}$ particle carries $8$ degrees of freedom, thus the remaining $4$ d.o.f. are described by a $j=\tfrac{1}{2}$ particle. In a generic Lorentz frame the two spin states can be singled out by applying spin projectors, with the $j=\tfrac{3}{2}$ particle defined by the transversality condition $p^\mu \psi_\mu =0$.

In what follows we explicitly construct the coefficient functions for a generic vector-spinor $\psi_\mu$ subject to Eq.\ \eqref{eq:vector-spinor-irrep} such that the following causality condition is satisfied:
\begin{equation}\label{eq:causality-condition}
	\{ \psi_\mu (x) , \psi^\dagger_\nu (y) \} = 0 \, \quad{\rm for} \quad (x-y)^2>0 \, .
\end{equation}
Let us commence by writing down the ansatz of the mode decomposition of the gamma-traceless field $\psi_\mu$, i.e.,
\begin{equation}\label{eq:first-mode-expansion-psi}
	\begin{split}
	\psi_\mu (x) =& \, \psi^{+}_\mu (x) + \psi^{-}_\mu(x) \\
	= & \, \sum_{\alpha=\tfrac{1}{2},\tfrac{3}{2}} \sum_s \int \frac{d^3 p}{(2\pi)^{3/2}} \big[ \varphi^\alpha_\mu (\boldsymbol{p},s) e^{i p \cdot x} a_\alpha (\boldsymbol{p},s) \\
	&  +  \chi^\alpha_\mu (\boldsymbol{p},s) e^{-i p \cdot x} a_\alpha^{c\, \dagger} (\boldsymbol{p},s) \big] \, ,
	\end{split}
\end{equation}
where $\psi^{+}_\mu$ and $\psi^{-}_\mu$ are the annihilation and creation operators. Moreover, $\alpha$ is summed over the two irreducible representations of rotations, i.e., $\alpha=\tfrac{1}{2},\,\tfrac{3}{2}$. Thus, the only nonvanishing anticommutation relation are
\begin{align}\label{eq:anticommutation-first}
 	\{ a_\alpha (\boldsymbol{p},s) , a_{\alpha'}^\dagger (\boldsymbol{p}',s') \} = \delta^3(\boldsymbol{p}-\boldsymbol{p}') \delta_{ss'} \delta_{\alpha \alpha'} \, .
 \end{align}
Notice that the coefficient functions are dimensionless.

Consider a matrix $\boldsymbol{J}^{(j)}$ providing an irreducible spin-$j$ representation of the rotation group, and let ${\cal J}_I{}^{I'}$ be the generators of rotations. Then, in the rest frame the coefficient functions $\Phi_I(\boldsymbol{0},s)$ and $\Xi_I(\boldsymbol{0},s)$ of the annihilation and creation operators must satisfy, respectively \cite{Weinberg:1964cn,Weinberg:1995mt},
\begin{subequations}
	\begin{align}\label{eq:spin-and-rotations}
		& \sum_{s'}\left(\boldsymbol{J}^{(j)}\right)_{ss'} \Phi_I(\boldsymbol{0},s') = {\cal J}_I{}^{I'}\, \Phi_{I'} (\boldsymbol{0},s) \, ;\\
		& - \sum_{s'}\left(\boldsymbol{J}^{*\,(j)}\right)_{ss'} \Xi_I(\boldsymbol{0},s') = {\cal J}_I{}^{I'}\, \Xi_{I'} (\boldsymbol{0},s) \, ,
	\end{align}
\end{subequations}
where $s,s'=j, \dots , -j \,$ and $*$ stands for complex conjugation. These equations represent the starting point for the derivation of the coefficient functions of any massive causal quantum field.

A vector-spinor $\psi_\mu$ is described in terms of its temporal and spatial parts as
\begin{equation}\label{eq:time-space-splitting-psi}
	\psi_\mu = \left( \psi_i,\psi_0 \right),
\end{equation}
and likewise for the coefficient functions $\varphi_\mu$ and $\chi_\mu$ of the annihilation and construction operators. While the temporal component can only transform as a $j=\tfrac{1}{2}$ Dirac spinor under rotations, the spatial ones are given by a sum of $j=\tfrac{3}{2}$ and $j=\tfrac{1}{2}$ states. To better grasp this feature, let us write down the generators of rotation that act on these spaces. In the vector and Dirac spinor representation a rotation around the $k$ direction takes the following simple forms
\begin{eqnarray}\label{eq:rotations-vector-and-spinor}
	({\cal J}_k)^{{\rm v}\,i}{}_j = -i \epsilon^i{}_{jk} \, , \quad && ({\cal J}_k)^{\rm s} = \frac{1}{2} 
	\begin{pmatrix}
		\sigma_k & 0\\
		0 & \sigma_k 
	\end{pmatrix} \, ,
\end{eqnarray}
where $\sigma_i$ are the Pauli matrices. Thus, in the vector-spinor representation we have
\begin{equation}\label{eq:vector-spinor-rep-squared}
	 ({\cal J}_k)^{{\rm vs} \,i}{}_{j} = -i \epsilon^i{}_{jk} \id + \frac{1}{2} {\small \begin{pmatrix}
	 	\sigma_k & 0\\
	 	0 & \sigma_k 
	 \end{pmatrix}} \delta^i{}_j \, .
\end{equation}
Let us look at Eq.\ \eqref{eq:spin-and-rotations}. By squaring it we have that the left-hand side is diagonal, and it reads $j(j+1) \delta_{ss'} \Phi_I(\boldsymbol{0},s)$. Instead, on the spatial components $\psi_i$ on the right-hand side we are acting with
 \begin{equation}
 	({\cal J}_k)^{{\rm vs} \,i}{}_{l} ({\cal J}_k)^{{\rm vs} \,l}{}_{j}  = \frac{15}{4} \delta^i{}_j \id -  {\small \begin{pmatrix}
 			\sigma^i \sigma_j & 0\\
 			0 & \sigma^i \sigma_j 
 	\end{pmatrix}} \, . 
 \end{equation}
Thus, the spin $j=\tfrac{3}{2}$ part of $\psi_i$ is that whose coefficient functions satisfy
\begin{equation}\label{eq:spin-3/2-condition}
	{\small \begin{pmatrix}
			\sigma^i & 0\\
			0 & \sigma^i 
	\end{pmatrix}} \varphi^{3/2}_i (\boldsymbol{0},s) = 0 \, , \,\,\, {\small \begin{pmatrix}
	\sigma^i & 0\\
	0 & \sigma^i 
\end{pmatrix}} \chi^{3/2}_i (\boldsymbol{0},s) = 0 \, ,
\end{equation}
while we have a spin $j=\tfrac{1}{2}$ state provided that
\begin{equation}\label{eq:spin-1/2-condition}
	\begin{split}
	&{\small \begin{pmatrix}
			\sigma^i \sigma_j & 0\\
			0 & \sigma^i \sigma_j
	\end{pmatrix}} \varphi^{1/2}_j (\boldsymbol{0},s) = 3 \, \varphi^{1/2}_i (\boldsymbol{0},s) \, , \\
	& {\small \begin{pmatrix}
			\sigma^i \sigma_j & 0\\
			0 & \sigma^i \sigma_j
	\end{pmatrix}} \chi^{1/2}_j (\boldsymbol{0},s) = 3 \, \chi^{1/2}_i (\boldsymbol{0},s) \, .
	\end{split}
\end{equation}
From the last two equations we see that a possible parametrization of the $j=\tfrac{1}{2}$ part of $\varphi_i$ and $\chi_i$ is given by
\begin{equation}
	\begin{split}
	& \varphi^{1/2}_j (\boldsymbol{0},s) \propto \tfrac{1}{3} \gamma_i \, u(\boldsymbol{0},s) \, ;\\
	& \chi^{1/2}_j (\boldsymbol{0},s) \propto \tfrac{1}{3} \gamma_i \, v(\boldsymbol{0},s) \, ,
\end{split}
\end{equation}
where $u$ and $v$ are the zero-momentum coefficient functions of the annihilation and creation operators of a Dirac field. Let us also observe that the relative phases of the left- and right-handed spinor subspaces cannot be fixed at this level, hence a $\gamma_5$ factor may appear on the rhs of the previous equations.

As we have said above, there is a second $j=\frac{1}{2}$ mode, which is carried by $\psi_0$. However, the gamma-traceless condition Eq.\ \eqref{eq:vector-spinor-irrep} implies that only a linear combination of the two $j=\tfrac{1}{2}$ modes is physical. Indeed, a covariant way to describe the $j=\tfrac{1}{2}$ particle carried by the vector-spinor is provided by
\begin{equation}\label{eq:spin-1/2-parametrization}
	\psi^{1/2}_\mu = \left( \partial_\mu - \tfrac{1}{4} \gamma_\mu \fsl{\partial}\right) \lambda \, ,
\end{equation}
where $\lambda$ is a Dirac spinor and the tensor structure on the rhs is dictated by the gamma-traceless condition. In the rest frame this boils down to
\begin{equation}
	\begin{split}
	& \varphi^{1/2}_{0}(\boldsymbol{0},s) \propto - \frac{3i}{4} p^0 u (\boldsymbol{0},s) \, , \\
	&\varphi^{1/2}_i (\boldsymbol{0},s) \propto - \frac{i}{4} p^0 \begin{pmatrix}
		 \sigma_i & 0 \\
		 0 & -\sigma_i 
	\end{pmatrix} u (\boldsymbol{0},s) \, ,
\end{split}
\end{equation}
where $s=\pm\tfrac{1}{2}$ and similar equations hold for $\chi^{1/2}_\mu(\boldsymbol{0},s)$ too. Notice that the overall normalization must be determined relative to that of the $j=\tfrac{3}{2}$ coefficient functions. 

We now derive the specific form of the coefficient functions, starting from the spin $j=\tfrac{3}{2}$ part. In this case the generators of rotations are
\begin{equation}
 	J^{3/2}_+ = 
{\small\begin{bmatrix}
	0 & \sqrt{3} & 0 & 0\\
	0 & 0 & 2 & 0 \\
	0 & 0 & 0 & \sqrt{3} \\
	0 & 0 & 0 & 0
\end{bmatrix} }\, , \quad
	\left(J^{3/2}_3 \right)_{s s'} = s \, \delta_{ss'} ,
\end{equation}
where the raising and lowering matrices are defined by
\begin{equation}
	J_{\pm} \equiv J_1 \pm i J_2 \, ,
\end{equation}
and $J_{-}=\left(J_{+}\right)^t$. Choosing the $3$-direction in Eq.\ \eqref{eq:spin-and-rotations} and acting on $\varphi^{3/2}_{3}(\boldsymbol{0},s= \pm \frac{3}{2})$ we derive
\begin{equation}
 \varphi^{3/2}_{3}(\boldsymbol{0},s= \pm \frac{3}{2}) = 0 \, .
\end{equation}
Furthermore, Eq.\ \eqref{eq:spin-3/2-condition} restricts the other two spatial components' relative coefficients, yielding
\begin{equation}
	\varphi^{3/2}_{\mu}(\boldsymbol{0},\tfrac{3}{2}) = \frac{1}{2} \begin{pmatrix}
		{\small\begin{pmatrix}
			b_{+}\\
			0\\
			b_{-}\\
			0
		\end{pmatrix}} \, , \,  {\small\begin{pmatrix}
		i b_{+}\\
		0\\
		i b_{-}\\
		0
		\end{pmatrix}} \, , 0 \, , 0 
	\end{pmatrix}^t \, ,
\end{equation}
One of the constants $b_{\pm}$ can always be chosen by specifying the overall normalization of the field, obtaining
\begin{equation}
	\varphi^{3/2}_{\mu}(\boldsymbol{0},\tfrac{3}{2}) = \frac{1}{2} \begin{pmatrix}
		{\small\begin{pmatrix}
			1\\
			0\\
			b\\
			0
		\end{pmatrix}} \, , \,  {\small\begin{pmatrix}
			i \\
			0\\
			i b\\
			0
		\end{pmatrix}} \, , 0 \, , 0 
	\end{pmatrix}^t \, .
\end{equation}
The remaining spin eigenstates of $\varphi^{3/2}$ are determined by repeatedly using Eqs.\ \eqref{eq:spin-and-rotations} \eqref{eq:spin-3/2-condition}, and the results are collected in Eqs.\ \eqref{eq:varphi-3/2-app-1} and \eqref{eq:varphi-3/2-app-2}. Using the same strategy we work out the coefficient functions of the destruction operators. This time a different constant $c$ enters the result, see Eq.\ \eqref{eq:chi-3/2-app}. By requiring that the field transforms in a simple fashion under parity we deduce that these constants must be sign factors \cite{Weinberg:1995mt}.

Having derived the parametric form of the coefficient functions in the rest frame, we can compute the spin sums. After some algebra we find
\begin{equation}\label{eq:spin-sums-rest-frame}
	\begin{split}
		&\sum_s \varphi^{3/2}_{i}(\boldsymbol{0},s) \varphi^{3/2}_{j}{}^\dagger(\boldsymbol{0},s) = \frac{\left( \id + b \beta \right)}{2} \left( \id \delta_{ij} - \frac{1}{3} \gamma_i \gamma_j \right) \, ,\\
		&\sum_s \chi^{3/2}_{i}(\boldsymbol{0},s) \chi^{3/2}_{j}{}^\dagger(\boldsymbol{0},s) = \frac{\left( \id + c \beta \right)}{2} \left( \id \delta_{ij} - \frac{1}{3} \gamma_i \gamma_j \right) \, ,
	\end{split}
\end{equation}
where $\beta\equiv i \gamma^0$. Notice that both equations are consistently $\gamma_i$-traceless on the left- and right-hand sides. Moreover, the spin sums are clearly orthogonal to the time direction. Thus, in a generic Lorentz frame the spin sums must be orthogonal to both $p_\mu$ and $\gamma_\mu$. Hence, they are given by
\begin{equation}\label{eq:spin-sums}
	\begin{split}
 	\sum_s \varphi^{3/2}_{\mu}(\boldsymbol{p},s) \varphi^{3/2}_{\nu}{}^\dagger(\boldsymbol{p},s) = & \frac{1}{2p^0} \left( -i \fsl{p} + b \, m_{3/2} \right)\\
 	& \times \left(P^{3/2}\right)_{\mu\nu} \beta \, , \\
 	\sum_s \chi^{3/2}_{\mu}(\boldsymbol{p},s) \chi^{3/2}_{\nu}{}^\dagger(\boldsymbol{p},s) = & \frac{1}{2p^0} \left( -i \fsl{p} + c \, m_{3/2} \right) \\
 	& \times \left(P^{3/2}\right)_{\mu\nu} \beta \, ,
	\end{split}
\end{equation}
where the result for the Dirac case \cite{Weinberg:1995mt} simply multiplies the $j=\tfrac{3}{2}$ projector.

Now we focus on the spin $j=\tfrac{1}{2}$ part of the vector-spinor. Let $u(\boldsymbol{0},s)$ and $v(\boldsymbol{0},s)$ be the zero-momentum coefficient functions of the annihilation and creation operators. They can be written in terms of a scaling constant $\kappa$ and two sign factors $f$ and $g$ as
{\small\begin{subequations}
 \begin{align}
 	& u(\boldsymbol{0},\tfrac{1}{2}) = \kappa \begin{pmatrix}
 		1\\
 		0\\
 		f\\
 		0\\
 	\end{pmatrix} \, , \qquad  u(\boldsymbol{0},-\tfrac{1}{2}) = \kappa \begin{pmatrix}
 	0\\
 	1\\
 	0\\
 	f\\
 	\end{pmatrix} \, ;\\
 	& v(\boldsymbol{0},\tfrac{1}{2}) = \kappa \begin{pmatrix}
 		0\\
 		1\\
 		0\\
 		g\\
 	\end{pmatrix} \, , \qquad  v(\boldsymbol{0},-\tfrac{1}{2}) = \kappa \begin{pmatrix}
 		-1\\
 		0\\
 		-g\\
 		0\\
 	\end{pmatrix} \, .
 \end{align}
\end{subequations}}
Hence, the spin sums are
\begin{equation}
	\begin{split}
		& \sum_s u(\boldsymbol{p},s) u^\dagger (\boldsymbol{p},s) = \frac{\kappa^2}{2p^0} \left( -i p^\mu \gamma_\mu + f \, m_{1/2} \right) \beta \, ,\\
		& \sum_s v(\boldsymbol{p},s) v^\dagger (\boldsymbol{p},s) = \frac{\kappa^2}{2p^0} \left( -i p^\mu \gamma_\mu + g \, m_{1/2} \right) \beta \, .
	\end{split}
\end{equation}
The mode functions $\varphi^{1/2}_\mu$ and $\chi^{1/2}_\mu$ are found by using the explicit form of the longitudinal parametrization Eq.\ \eqref{eq:spin-1/2-parametrization}, yielding
\begin{subequations}
 \begin{align}
 	& \varphi^{1/2}_\mu (\boldsymbol{p},s) = i \left( p_\mu - \frac{1}{4} \gamma_\mu \fsl{p} \right) u (\boldsymbol{p},s) \, ,\\
 	& \chi^{1/2}_\mu (\boldsymbol{p},s) = - i \left( p_\mu - \frac{1}{4} \gamma_\mu \fsl{p} \right) v (\boldsymbol{p},s) \, .
 \end{align}
\end{subequations}
Accordingly, the spin sums of the $j=\tfrac{1}{2}$ annihilation coefficient functions can be rearranged as
\begin{align}\nonumber\label{eq:spin-sum-spin1/2-annihilation}
 	& \sum_s \varphi^{1/2}_\mu (\boldsymbol{p},s) \varphi^{1/2}_\nu{}^\dagger (\boldsymbol{p},s) = \, \left( p_\mu - \frac{1}{4} \gamma_\mu \fsl{p} \right) \frac{\kappa^2}{2p^0} \times \\\nonumber
 	& \times \left( - i \fsl{p} + f \, m_{1/2} \right) \beta \left( p_\nu - \frac{1}{4} \fsl{p} \gamma_\nu \right) \\
 	= & \, \frac{\kappa^2}{2p^0} \left( - 2 i \fsl{p}\delta_\mu{}^\rho + i \gamma_\mu p^\rho + f \, m_{1/2} \delta_\mu{}^\rho \right) \times \\\nonumber
 	& \times \left( p_\rho - \frac{1}{4} \gamma_\rho \fsl{p} \right) \left( p_\nu - \frac{1}{4} \fsl{p} \gamma_\nu \right) \beta \\\nonumber
 	= & \, \frac{3 \kappa^2 \, p^2}{p^0} \left( - i \fsl{p} \delta_\mu{}^\rho + \frac{i}{2} \gamma_\mu p^\rho + f \, \frac{m_{1/2}}{2} \delta_\mu{}^\rho \right)  \left(P^{\tfrac{1}{2}}_{BB}\right)_{\rho\nu} \beta \, .
\end{align}
A completely analogous calculation shows that
\begin{align}\label{eq:spin-sum-spin1/2-creation}
  &\sum_s \chi^{1/2}_\mu (\boldsymbol{p},s) \chi^{1/2}_\nu{}^\dagger (\boldsymbol{p},s) = \\\nonumber
  & = \, \frac{3 \kappa^2 \, p^2}{p^0} \left( - i \fsl{p} \delta_\mu{}^\rho + \frac{i}{2} \gamma_\mu p^\rho +  g \, \frac{m_{1/2}}{2} \delta_\mu{}^\rho \right)  \left(P^{\tfrac{1}{2}}_{BB}\right)_{\rho\nu} \beta \, .
\end{align}
Now use the fact that both the $j=\tfrac{3}{2}$ and $j=\tfrac{1}{2}$ particles are described by the single field variable $\psi_\mu$. By combining the spin sums of the annihilation coefficients we obtain
\begin{align}\label{eq:spin-sums-full-annihilation}\nonumber
	& \sum_{\alpha=\tfrac{1}{2},\tfrac{3}{2}} \sum_s \varphi^{\alpha}_\mu (\boldsymbol{p},s) \varphi^{\alpha}_\nu{}^\dagger (\boldsymbol{p},s) = \frac{1}{2p^0} \left[ \left( -i \fsl{p} \delta_\mu{}^\rho + b \, m_{3/2} \delta_\mu{}^\rho \right) \right.\\\nonumber
	& \left. \times \left( P^{\tfrac{3}{2}} \right){}_{\rho\nu} + 6 \kappa^2 p^2 \left( -i \fsl{p} \delta_\mu{}^\rho + \frac{i}{2} \gamma_\mu p^\rho + \frac{f}{2} m_{1/2} \delta_\mu{}^\rho \right) \right.\\
	& \left. \times \left( P^{\tfrac{1}{2}}_{BB} \right)_{\rho\nu} \right] \beta \, ,
\end{align}
with a similar equation holding for the construction coefficients. Since the two spin modes are to be described by the \emph{sole} field variable $\psi_\mu$, we must be able to factor out the gamma-traceless projector, say on the right. Then, using the definition of the gamma-traceless projector Eq.\ \eqref{eq:def-gamma-traceless-proj} we obtain
\begin{eqnarray}
	\kappa^2 = \frac{1}{6 p^2} \, , \qquad &&  b \, m_{1/2} = 2 f \, m_{3/2} \, .
\end{eqnarray}
Given that $b$ and $f$ are sign factors, these equations are solved by $\kappa=\pm \tfrac{1}{\sqrt{6|p^2|}}$, $b=f$ and $m_{1/2}=2m_{3/2}$, where we notice a sign ambiguity in the value of $\kappa$. The last equation correctly reproduces the mass ratio of the $j=\tfrac{3}{2}$ and $j=\tfrac{1}{2}$ modes obtained in the classical analysis of the previous section. The counterpart of Eq.\ \eqref{eq:spin-sums-full-annihilation} for the construction coefficients yields $c=g$. Then, as a direct consequence of these findings and using Eq.\ \eqref{eq:anticommutation-first}, the anticommutator of the field and its adjoint can be written as
\begin{align}\label{eq:anticommutation-second}
	\{ \psi_\mu (x),\psi_\nu^\dagger (y) \} = & \left( - \fsl{\partial} \delta_\mu{}^\rho + \frac{1}{2} \gamma_\mu \partial^\rho + b \, m_{3/2} \delta_\mu{}^\rho \right) \\\nonumber
	& \times \left( P^{\gamma T} \right)_{\rho\nu} \beta \Delta_{+}(x-y) \\\nonumber
	& + \left( - \fsl{\partial} \delta_\mu{}^\rho + \frac{1}{2} \gamma_\mu \partial^\rho + c \, m_{3/2} \delta_\mu{}^\rho \right) \\\nonumber
	& \times \left( P^{\gamma T} \right)_{\rho\nu} \beta \Delta_{+}(y-x) \, ,
\end{align}
where
\begin{equation}
	\Delta_{+} (x) \equiv \int \frac{d^3 p}{2p^0(2\pi)^3} e^{i \, p\cdot x} \, .
\end{equation}
Since this function is even when its argument is spacelike \cite{Weinberg:1995mt}, causality implies that
\begin{equation}
	c = -b \, .
\end{equation}
Finally, up to a redefinition of $\psi_\mu$ involving $\gamma_5$ we can take $b=1$, and we are now free to drop the subscript in the masses, choosing $m=m_{3/2}$. As a consequence of the present construction and exploiting the transversality $p^\mu \varphi^{3/2}_\mu=p^\mu \chi^{3/2}_\mu = 0$, we observe that the coefficient functions satisfy the following equations
\begin{equation}\label{eq:field-equation-coefficient-functions}
	\begin{split}
		& \left( i \fsl{p} \delta_\mu{}^\nu - \frac{i}{2} \gamma_\mu p^\nu + m \delta_\mu{}^\nu \right) \varphi^\alpha_\nu (\boldsymbol{p},s) = 0 \, ;  \\
		& \left( - i \fsl{p} \delta_\mu{}^\nu + \frac{i}{2} \gamma_\mu p^\nu + m \delta_\mu{}^\nu \right) \chi^\alpha_\nu (\boldsymbol{p},s) = 0 \, , 
	\end{split}
\end{equation}
with $\alpha=\tfrac{1}{2},\tfrac{3}{2}$. Thus, the outcome of imposing Poincaré invariance of the free field $\psi_\mu(x) \in \left( 1, \tfrac{1}{2} \right) \oplus \left( \tfrac{1}{2}, 1 \right)$ is that this field propagates two spin states according to the field equation
\begin{equation}
	\left(  \fsl{\partial} \, \delta_\mu{}^\nu - \frac{1}{2} \gamma_\mu \partial^\nu + m \, \delta_\mu{}^\nu \right) \psi_\nu (x) = 0 \, ,
\end{equation}
in perfect accordance with the \emph{ab initio} analysis performed in the previous section \ref{sect:ferm-gauge-transf} (see Eq.\ \eqref{eq:field-equations}).

The behavior of a vector-spinor under discrete transformations can be derived by combining the known results for vectors and Dirac spinors \cite{Weinberg:1995mt}. In particular, under parity we have
\begin{equation}
	P \psi_\mu (x) P^{-1} = - \eta^{*} \beta \psi_\mu ({\cal P} x) \, ,
\end{equation}
where $\eta$ is the intrinsic parity of the $j=\tfrac{3}{2}$ particles carried by $\psi_\mu$.
Antiparticles transform in a similar way, with $\eta^c$ replacing $\eta^*$. In particular, causality implies that $\eta^c \, \eta^{*}= -1$ like in the Dirac case. 

Finally, the sign ambiguity in the numerical value of $\kappa$ can be understood as a discrete $\mathbb{Z}_2$ symmetry in the $j=\tfrac{1}{2}$ sector. Thus, we may always let $\varphi^{1/2}_\mu \rightarrow - \varphi^{1/2}_\mu$, $\chi^{1/2}_\mu \rightarrow - \chi^{1/2}_\mu$.

\subsection{The propagator}\label{subsect:propagator}

One of the possible definitions of the propagator $\Delta_{\mu\nu}(x-y)$ (see \cite{Weinberg:1995mt}, Chapter $6$) is written in terms of the annihilation and creation operators $\psi^{+}_\mu$ and $\psi^{-}_\mu$ as
\begin{equation}\label{eq:def-propagator}
	\begin{split}
		- i \Delta_{\mu\nu} (x-y) \equiv & \,\theta(x-y) \{ \psi^{+}_\mu (x) , \psi^{+ \, \dagger}_\nu (y) \}\\
		& - \theta(y-x) \{ \psi^{-}_\nu (y) , \psi^{- \, \dagger}_\mu (x) \} \, .
	\end{split}
\end{equation}
Accordingly, the Feynman propagation is defined as
\begin{equation}
	\Delta_F(x) \equiv \theta(x) \Delta_+(x) + \theta(-x) \Delta_{+}(-x) \, ,
\end{equation}
where $\theta(x)$ is the Heaviside step function whose integral representation is given by
\begin{equation}
	\theta(t) = \frac{i}{2\pi} \int_{-\infty}^{\infty} d\omega \frac{e^{-i\omega t}}{\omega + i \epsilon} \, .
\end{equation}
By using the explicit results for the sums of the coefficient functions in Eq.\ \eqref{eq:spin-sums-full-annihilation} we then find the full expression of the propagator,
\begin{equation}\label{eq:feynman-prop-full}
\begin{split}
	- i \Delta_{\mu\nu} (x-y) = & - i \int \frac{d^4p}{(2\pi)^4} \frac{e^{ip\cdot (x-y)}}{p^2+m^2-i\epsilon} (P^{3/2})_\mu{}^\rho \times\\
	& \times \left( -i \fsl{p} + m \right) (P^{3/2})_{\rho\nu} \beta + \\
	& - i \int \frac{d^4p}{(2\pi)^4} \frac{e^{ip\cdot(x-y)}}{p^2+4m^2-i\epsilon} (P^{\gamma T})_{\mu\rho} \times\\
	& \times \frac{p^\rho p^\lambda}{p^2} \left( - i \fsl{p} + 2 m \right) (P^{\gamma T})_{\lambda\nu} \beta  \, ,
\end{split}
\end{equation}
where we notice the presence of two distinct contributions from the two separate spin eigenstates. While the $j=3/2$ mode behaves as expected, the lower-spin component does not. Indeed, we may define its propagator as by considering the divergences of the annihilation and creation operators in Eq.\ \eqref{eq:def-propagator}, i.e.,
\begin{equation}
	\begin{split}
		-i \Delta^{1/2} (x-y) & \equiv \theta(x-y) \{ \partial^{x,\mu} \psi^+_\mu(x), \partial^{y,\nu} \psi^{+ \, \dagger}_\nu (y) \} + \\
		& - \theta(y-x) \{ \partial^{y,\nu} \psi^{- \, \dagger}_\nu (y) ,\partial^{x,\mu} \psi^-_\mu(x) \} \, .
	\end{split}
\end{equation}
Then, the explicit computation of this expression shows that
\begin{equation}\label{eq:feynman-prop-spin-1/2}
	- i \Delta^{1/2}(x-y) = - \frac{3i}{32} \int \frac{d^4p}{(2\pi)^4} \frac{p^2 \left( - i \fsl{p} + 2m\right) \beta}{p^2 + 4 m^2 - i \epsilon} \, , 
\end{equation}
where we notice that the lower-spin mode displays a residue with the opposite sign with respect to ordinary Dirac particles at the pole $p^2 = - 4 m^2$, i.e., it is a ghost. Of course, all $\beta$ factors on the right-hand sides of the previous equations disappear if we take into account $\barpsi_\mu$ instead of $\psi^\dagger_\mu$.

According to the present analysis no difficulty arises in defining the Feynman propagator for a gamma-traceless vector-spinor $\psi_\mu$. Moreover, Eq.\ \eqref{eq:field-equation-coefficient-functions} shows that this propagator is consistent with the Lagrangian description of this field based on the gauge invariant action Eq.\ \eqref{eq:gauge-fixed-action}. However, the theory propagates a negative-norm state in the lower-spin sector.

\section{Quantum theory: the conformal anomaly}\label{sect:quantum}

A standard procedure of deriving the divergent part of the effective action on curved spacetime is to rely on the heat kernel technique \cite{Schwinger:1951nm,DeWitt:1964mxt,Vassilevich:2003xt}. To this end, one notices that the spectrum of the $\gamma_5$-adjoint of the linear differential operator $\Lambda_\mu{}^\nu$ and that of $\Lambda_\mu{}^\nu$ itself are identical, i.e.,
\begin{equation}
	{\rm Tr} \log \left( \Lambda_\mu{}^\nu \right) = {\rm Tr} \log \left( \gamma_5 \Lambda_\mu{}^\nu \gamma_5 \right) \, .
\end{equation}
Therefore, one can express the spectrum of the first-order differential operator $\Lambda_\mu{}^\nu$ in terms of the following second-order one
\begin{equation}\label{eq:def-second-order-operator-part1}
	\begin{split}
	{\rm Tr} \log \left( \Lambda_\mu{}^\nu \right) & = \frac{1}{2} {\rm Tr} \log \left( \Lambda_\rho{}^\mu \gamma_5 \Lambda_\mu{}^\nu \gamma_5 \right)\\
	& \equiv \frac{1}{2} {\rm Tr} \log \left( {\cal H}_\rho{}^\lambda \right) \, .
	\end{split}
\end{equation}
To compute the divergent part of the effective action we employ the heat kernel technique, which rests on the definition of the kernel \cite{DeWitt:1964mxt,Barvinsky:1985an}, i.e.,
\begin{equation}
	\hat{\mathit K} (x,y;s) \equiv e^{i s \hat{\cal H}(x)} \delta(x,y) \, ,
\end{equation}
where $s$ is the so-called proper time and has dimension of a squared length. From the definition, one derives the following Schr\"odinger-like equation for the kernel:
\begin{equation}
	\left(i \partial_s + \hat{\cal H}(x) \right) \hat{\mathit K} (x,y;s) = 0 \, .
\end{equation}
Then, the trace of the logarithm of the operator $\hat{\cal H}$ can be written as
\begin{equation}\label{eq:trace-log-asymptotics}
	\begin{split}
		{\rm Tr} \log \hat{\cal H} = & - \int_0^\infty \frac{ds}{s} {\rm Tr} \, e^{i s \hat{\cal H}} \equiv - \int_0^\infty \frac{ds}{s} {\rm Tr} \hat{\mathit K}  \\
		= & \sum_{n\geq0} \frac{i^{n+1} }{(4\pi i)^{d/2} } \int_0^\infty ds \, s^{n-1-\tfrac{d}{2}} \, {\rm tr} \, \hat{a}_n \, ,
	\end{split}
\end{equation}
where in the last step we have inserted the asymptotic expansion of the heat kernel valid for second-order operators \cite{DeWitt:1964mxt}. When the operator $\hat{\cal H}$ is nonminimal, i.e., its principal part is not just given by a box operator, the asymptotic expansion can be written only in the coincidence limit $y\rightarrow x$, which is sufficient to compute the divergent part of the effective action. The coefficients of this expansion are known as Seeley-DeWitt coefficients $\hat{a}_n$, and they are local covariant functions of the dimensionful tensors that describe the operator \cite{Barvinsky:1985an}. Dimension-four counterterms are then extracted by computing the $\hat{a}_2$ coefficient, which is also proportional to the conformal anomaly \cite{Vassilevich:2003xt}. Thus, the computation of this coefficient will be our goal henceforth.

The second-order operator defined by Eq.\ \eqref{eq:def-second-order-operator-part1} can be written in a convenient way by retaining the gamma-traceless projectors on the left and on the right, thus defining
\begin{align}\label{eq:def-second-order-operator}\nonumber
		{\cal H}_\rho{}^\lambda = &\, (P^{\gamma T})_\rho{}^\mu \, \left( \fsl{\nabla} + m \right) \, (P^{\gamma T})_\mu{}^\sigma \, \left( \fsl{\nabla} - m \right) \, (P^{\gamma T})_\sigma{}^\lambda\\
		\equiv & \, (P^{\gamma T})_\rho{}^\mu \, \tilde{{\cal H}}_\mu{}^\nu \, (P^{\gamma T})_\nu{}^\lambda \, .
\end{align}
The explicit form of $\tilde{{\cal H}}_\mu{}^\nu$ in $d=4$ is given by
\begin{align}\label{eq:second-order-op-reduced}
	\tilde{{\cal H}}_\mu{}^\nu = & \,\square \, \delta_\mu{}^\nu - \nabla_\mu \nabla^\nu - \frac{1}{4} R \, \delta_\mu{}^\nu - \frac{1}{2} R^\nu{}_{\mu\alpha\beta} \, \gamma^{\alpha\beta} \\\nonumber
	& - m^2 \delta_\mu{}^\nu \, .
\end{align}
Let us make some remarks on some crucial properties of this nonminimal second-order operator. Since the explicit parametrization of longitudinal modes is proportional to $\nabla_\mu \lambda - \frac{1}{4} \gamma_\mu \fsl{\nabla} \lambda$, they do not belong to the kernel of Eq.\ \eqref{eq:second-order-op-reduced}. Therefore, this second-order nonminimal operator has a nondegenerate principal part, and it belongs to the family of causal operators which have been first studied in \cite{Barvinsky:1985an}. Recently, using a refined version of the algorithm used in \cite{Melichev:2025hcg}, a model-independent result for the local terms of the divergent part of the effective action was found in \cite{Sauro:2025sbt}. This paper was published alongside an open-source $\mathtt{Mathematica}$ code that has been employed to carry out most of the subsequent computations.

To apply the heat kernel technique to nonminimal second-order operators one needs to rely on a trick that was first exploited in \cite{Barvinsky:1985an}. The idea is that the principal part of the operator $\tilde{{\cal H}}_\mu{}^\nu$ can be written as a special case of a one-parameter family of nonsingular second-order operators as
\begin{equation}\label{eq:principal-part}
	\begin{split}
	\tilde{\cal D}_\mu{}^\nu (\nabla) & = \square \delta_\mu{}^\nu - \nabla_\mu \nabla^\nu = \left( \square \delta_\mu{}^\nu - \zeta \nabla_\mu \nabla^\nu \right) \big|_{\zeta=1}\\
	& \equiv \tilde{\cal D}_\mu{}^\nu (\nabla;\zeta) \big|_{\zeta=1} \, .
	\end{split}
\end{equation}
Then, the full operator is simply given by the sum of the principal part and the rest as
\begin{equation}\label{eq:D-Y-splitting}
	\tilde{\cal H}_\mu{}^\nu = \tilde{\cal D}_\mu{}^\nu (\nabla;\zeta) \big|_{\zeta=1} + \tilde{Y}_\mu{}^\nu \, .
\end{equation}
The pseudoinverse of the principal part Eq.\ \eqref{eq:principal-part} is defined in momentum space (i.e.,\ by letting $\nabla_\mu \rightarrow n_\mu$)
\begin{equation}\label{eq:definition-K}
	\tilde{\cal D}_\mu{}^\rho (n) \tilde{\cal K}_\rho{}^\nu (n) = n^4 \delta_\mu{}^\nu \, .
\end{equation}
To go back to coordinate space we must choose an ordering of the covariant derivatives in the nonminimal term of $\tilde{\cal K}_\mu{}^\rho(\nabla)$. Our choice is given by
\begin{align}\label{eq:K-expression}
	\tilde{\cal K}_\rho{}^\nu (\nabla) = & \, \square \,\delta_\rho{}^\nu + \frac{4 \zeta}{4 - 3 \zeta} \nabla_\rho \nabla^\nu \equiv \, \tilde{\cal K}^{\alpha\beta}{}_\mu{}^\nu \nabla_\alpha\nabla_\beta \\\nonumber
	= & \left( g^{\alpha\beta} \delta_\mu{}^\nu + \frac{4 \zeta}{4-3\zeta} \delta^\alpha{}_\mu g^{\beta\nu} \right) \nabla_\alpha \nabla_\beta \, .
\end{align}
Furthermore, the nonminimal part of the differential operator Eq.\ \eqref{eq:principal-part} can be isolated 
\begin{equation}
	\tilde{\cal N}_\mu{}^\nu (\nabla; \zeta) \equiv - \zeta \nabla_\mu \nabla^\nu = \zeta \frac{d}{d\zeta} \tilde{\cal N}_\mu{}^\nu (\nabla; \zeta) \, .
\end{equation}
Consequently, by employing Eq.\ \eqref{eq:principal-part} we can write the trace log of the operator ${\cal H}_\mu{}^\nu$ as 
\begin{align}\label{eq:trace-log-splitting}
	{\rm Tr} \log {\cal H}_\mu{}^\nu = & \, {\rm Tr} \log \left( P^{\gamma T} \cdot \tilde{\cal D} \cdot P^{\gamma T}  \right)_\mu{}^\nu \quad \\\nonumber
	& + {\rm Tr} \log \left( P^{\gamma T} \cdot \tilde{\cal K} \cdot P^{\gamma T} \cdot \tilde{Y} \right)_\mu{}^\nu \\\nonumber
	= & \, {\rm Tr} \log \left( P^{\gamma T} \cdot \tilde{\cal D} (\zeta) \cdot P^{\gamma T}  \right)_\mu{}^\nu \big|_{\zeta=1} \\\nonumber
	& + {\rm Tr} \log \left( P^{\gamma T} \cdot \tilde{\cal K} \cdot P^{\gamma T} \cdot \tilde{Y} \right)_\mu{}^\nu\\\nonumber
	= & \, \int_0^1 \frac{d\zeta}{\zeta} {\rm Tr} \left( P^{\gamma T} \cdot \tilde{\cal N} \cdot P^{\gamma T} \cdot \tilde{\cal K} \cdot P^{\gamma T} \right)_\mu{}^\nu\\\nonumber
	& + {\rm Tr} \log \left( P^{\gamma T} \cdot \tilde{\cal K} \cdot P^{\gamma T} \cdot \tilde{Y} \right)_\mu{}^\nu \big|_{\zeta=1} \, .
\end{align}
Then, to work out the dimension-four divergent part of the effective action we are required to perform a perturbative expansion in the dimensionful operators and to evaluate the so-called universal functional traces $\nabla_{\mu_1} \dots \nabla_{\mu_p} \frac{1}{\,\,\,\square^n}$ \cite{Barvinsky:1985an,Sauro:2025sbt}. The first such dimensionful operator is found by commuting covariant derivatives in the coordinate-space version of Eq.\ \eqref{eq:definition-K}, which yields
\begin{align}\label{eq:D-tilde}
	\tilde{\cal D}_\mu{}^\rho (\nabla) \tilde{\cal K}_\rho{}^\nu (\nabla) \equiv \, & \square^2 \delta_\mu{}^\nu + \tilde{\cal M}_\mu{}^\nu (\nabla) \\\nonumber
	= \, & \square^2 \delta_\mu{}^\nu + \tilde{\cal M}(2)^{\alpha\beta}{}_\mu{}^\nu \nabla_\alpha \nabla_\beta + \dots  \, ,
\end{align}
where $2$ stands for the background dimension of the tensor coefficients $\tilde{\cal M}_\mu{}^\nu$ that are reported in Eq.\ \eqref{eq:M2-tilde}. The full expressions of the ${\cal K}^{\alpha\beta}{}_\rho{}^\lambda$ and ${\cal M}(2)^{\alpha\beta}{}_\rho{}^\lambda$ operators are found by contraction with gamma-traceless projectors, i.e.,
\begin{equation}\label{eq:K-and-M-to-tildeK-and-tildeM}
	\begin{split}
		& {\cal K}^{\alpha\beta}{}_\rho{}^\lambda = (P^{\gamma T})_\rho{}^\mu \, \tilde{\cal K}^{\alpha\beta}{}_\mu{}^\nu \, (P^{\gamma T})_\nu{}^\lambda \, ,\\
		& {\cal M}(2)^{\alpha\beta}{}_\rho{}^\lambda = (P^{\gamma T})_\rho{}^\mu \, \tilde{\cal M}(2)^{\alpha\beta}{}_\mu{}^\nu \, (P^{\gamma T})_\nu{}^\lambda \, .
	\end{split}
\end{equation}
The other two dimensionful operators that concur to the final form of the divergent part of the effective action are the generalized curvature and the endomorphism $\tilde{Y}_\mu{}^\nu$. The former is defined by
\begin{align}\label{eq:generalized-curvature-def}
	\Omega_{\mu\nu\,\rho}{}^\lambda \equiv \, (P^{\gamma T})_\rho{}^\sigma \, \frac{ \delta}{\delta \psi_\alpha} [\nabla_\mu ,\nabla_\nu] \psi_\sigma \, (P^{\gamma T})_\alpha{}^\lambda \, ,
\end{align}
while the latter is read off from the operator and Eq.\ \eqref{eq:D-Y-splitting}. The explicit expressions of these tensors are written in the Appendix, see Eqs.\ \eqref{eq:generalized-curvature} and \eqref{eq:endomorphism}.

To combine these tensor coefficients into the general equations derived in \cite{Sauro:2025sbt} it is particularly convenient to make use of a given basis in the Clifford algebra. This is done by splitting these tensors according to the Clifford basis as
\begin{align}\label{eq:clifford-split-M2}
	{\cal M}(2)^{\alpha\beta}{}_\rho{}^\lambda = & \, M(2)_s{}^{\alpha\beta}{}_\rho{}^\lambda \, \id + M(2)_{ps}^{\alpha\beta}{}_\rho{}^\lambda \, \gamma_5 \\\nonumber
	& + M(2)_{t}^{\alpha\beta}{}_\rho{}^\lambda{}^{\mu\nu} \, \gamma_{\mu\nu} \, ,
\end{align}
and likewise for the other tensors, where we have defined
\begin{subequations}
	\begin{align}
		& M(2)_s{}^{\alpha\beta}{}_\rho{}^\lambda \equiv \frac{1}{4} {\rm tr} \left( {\cal M}(2)^{\alpha\beta}{}_\rho{}^\lambda \right) \, ,\\
		& M(2)_{ps}{}^{\alpha\beta}{}_\rho{}^\lambda \equiv \frac{1}{4} {\rm tr} \left( {\cal M}(2)^{\alpha\beta}{}_\rho{}^\lambda \, \gamma_5 \right) \, ,\\
		& M(2)_t{}^{\alpha\beta}{}_\rho{}^\lambda{}^{\mu\nu} \equiv - \frac{1}{8} {\rm tr} \left( {\cal M}(2)^{\alpha\beta}{}_\rho{}^\lambda \, \gamma^{\mu\nu} \right) \, .
	\end{align}
\end{subequations}
The subscripts in the previous equations stand for scalar, pseudoscalar and tensor. This splitting enables us to write the general equations of \cite{Sauro:2025sbt} in a reasonably tractable manner, and most importantly it hugely simplifies the computation of the Dirac traces while keeping track only implicitly of the actual functional dependence in the tensor coefficients.

By combining the contributions coming from the two trace logs in Eq.\ \eqref{eq:trace-log-splitting} and computing the $\zeta$ integral we find the following result for the traced second heat kernel coefficient:
\begin{equation}\label{eq:traced-heat-kernel-coefficient}
	{\rm tr} \, a_2 = \frac{7}{20} W^2 + \frac{31}{120} E_4 + 4 m^2 R + 36 m^4 \, .
\end{equation}
We have chosen to express the higher-derivative part of this expression in terms of the so-called Weyl basis, which is given by the square of the Weyl tensor $W_{\mu\nu\rho\sigma}$, the Euler density $E_4$, and the square of the Ricci scalar (see, e.g., \cite{Percacci:2017fkn}). In the massless limit the starting Lagrangian \eqref{eq:gauge-inv-linear-differential-operator} is Weyl invariant, therefore, the traced second Seeley-DeWitt coefficient does not depend on the square of the Ricci scalar, whose infinitesimal Weyl transformation law is affine. 

This calculation yields a new expression for the conformal anomaly of gamma-traceless vector-spinors $\psi_\mu$ in the limit $m\rightarrow0$, which can be contrasted with the previously known result by Endo \cite{Endo:1994yj}, which is valid on Einstein spaces. Indeed, classically Weyl invariant actions generate the following local contributions to the conformal anomaly \cite{Komargodski:2011vj}:
\begin{equation}
	\langle T^\mu{}_\mu \rangle = \frac{1}{16\pi^2} \left( c \, W^2 - a E_4 \right) \, ,
\end{equation}
where the $a$ and $c$ coefficients are given by \cite{Komargodski:2011vj,Paci:2024xxz}
\begin{equation}
	\begin{split}
		a = & \frac{1}{360} \left( N_0 + 11 N_{1/2} + 62 N_1 - 93 N_{vs} \right)\\
		c = & \frac{1}{120} \left( N_0 + 6 N_{1/2} + 12 N_1 + 42 N_{vs} \right) \, .
	\end{split}
\end{equation}
Here, the numerical subscripts stand for the spin of the particles propagated by fields. Since in the present case the analysis of the propagating degrees of freedom has shown that we have more than one physical spin mode, we have opted for labeling the contribution of $\psi_\mu$ in a different manner. We observe that the contribution due to vector-spinors to the $a$ charge of the conformal anomaly is negative. This is in accordance with the general results of \cite{Hofman:2008ar,Komargodski:2011vj}, since this feature must be displayed whenever negative-norm states are propagating. Indeed, this is the case for the present theory, as it was noted by computing the propagator of the lower-spin component in Eq.\ \eqref{eq:feynman-prop-spin-1/2}. In passing we note that the contribution to the $a$ charge can be counterbalanced to yield zero if we consider a model involving nine scalars, two Dirac fermions, and one gauge field together with a vector-spinor. Finally, the total derivative part of the conformal anomaly cannot be computed with the present method, however, a newly found result \cite{Barvinsky:2025jbw} can be applied to work it out.

\section{Conclusions}\label{sect:outro}

Poincaré invariance and the cluster decomposition principle necessarily imply the necessity of a field description of particles \cite{Weinberg:1995mt}. However, the Rarita-Schwinger action fails to provide a consistent theory of spin-$\tfrac{3}{2}$ particles through a vector-spinor field as soon as interactions are introduced \cite{Johnson:1960vt,Velo:1969bt}. Back in $1964$ Weinberg's take on this issue was that it represents a drawback of the Lagrangian description of the field theory \cite{Weinberg:1964cn}. However, the Lagrangian formulation of Quantum Field Thoery (QFT) has proved to be crucial in the derivation and mathematical description of the standard model of particle physics. Thus, serious attempts should be made to retain the Lagrangian description in any tentative interacting theory of higher-spin particles. To this end, this problem may be stated in a different manner by saying that it is not possible to describe spin-$\tfrac{3}{2}$ particles through a consistent QFT that propagates this particle \emph{only}. Accordingly, the aim of the present paper was to show that a quantum theory of vector-spinors compatible with causality is found once this assumption is amended; nevertheless, the price to pay is the lack of unitarity \cite{Anselmi:1999bu,Anselmi:2020opi}. Therefore, future works should explore the possibility of coupling the theory to a Dirac field and introducing a further gauge invariance such that the lower-spin components are compatible with \emph{both} causality and unitarity.

This work provides a first attempt to find a solution to the long-standing conundrum of describing interacting higher-spin particles by focusing on the simplest example of $j=\tfrac{3}{2}$. The Lagrangian description is fixed by requiring the presence of an off-shell fermionic gauge symmetry valid regardless of the configurations of the background electromagnetic and gravitational fields. The resulting Lagrangian is unique, and it is given by a ``singular" element of the one-parameter family of operators describing spin-$\tfrac{3}{2}$ particles found in the literature \cite{Haberzettl:1998rw,Anselmi:1999bu}. The arguments that were proposed to discard such a theory are shown to be flawed in that they do not account for the gauge invariance of the action. Moreover, in the massless limit and in \emph{any} spacetime dimension this action fixes the arbitrary parameter that enters the most general Weyl invariant action for a vector-spinor.

By employing spin projectors, it is shown that the gamma-trace configurations $\gamma_\mu \chi$ of a generic vector-spinor $\Psi_\mu$ belong to the kernel of the action, and that they can be gauged away in a \emph{global} fashion. Thus, the theory involves only the irreducible representation of the Lorentz group given by $\psi_\mu \in \left( 1,\tfrac{1}{2} \right) \oplus \left( \tfrac{1}{2},1 \right)$, where $\gamma^\mu \psi_\mu =0$. The field equations $R^\mu=0$ of the theory are derived, and it is shown that only the secondary constraint $\nabla \cdot R = 0$ is present, which in the free field limit fixes the relative values of the masses of the $j=\tfrac{3}{2}$ and $j=\tfrac{1}{2}$ particles. Finally, the Velo-Zwanziger instability is proved to be absent.

A complementary analysis of the mode decomposition of a free causal quantum vector-spinor $\psi_\mu \in \left( 1,\tfrac{1}{2} \right) \oplus \left( \tfrac{1}{2},1 \right)$ is then carried out. The relative normalization of the two spin sectors is derived, and the mass of the spin-$\tfrac{1}{2}$ is found to be twice that of the spin-$\tfrac{3}{2}$ particle, in agreement with the previous result. Thus, the same field equations that are found by imposing the off-shell fermionic gauge invariance are obtained as eigenvalue equations for the coefficient functions, and the Feynman propagator is constructed. However, the latter shows that the lower-spin component gives rise to negative-norm states, thus hindering the unitarity of the theory.

Finally, the $1$-loop quantum fluctuations of the vector-spinor $\psi_\mu$ are considered using the heat kernel technique. The conformal anomaly is derived up to boundary terms, and the negative sign of the $a$-anomaly is noticed to be coherent with the general result of \cite{Hofman:2008ar,Komargodski:2011vj}.

Future works are needed to understand if the present picture can be extended also to fermionic tensor fields of higher rank, generalizing the flat-space results of \cite{Anselmi:1999bu}. In particular, it is compelling to know if the consistency between the field theoretical description based on an off-shell fermionic gauge invariance and the mode decomposition of a causal quantum field belonging to an irreducible representation of the Lorentz group is valid beyond the $j=\tfrac{3}{2}$ case. Furthermore, it is natural to ask ourselves if any such extended theory is also Weyl invariant. Finally, the physically interesting aspects of this field-theoretical work will be to comprehend how many particles are propagated by a rank-$k$ totally symmetric fermionic field, which are the ratios between the masses, and foremost if it is possible to couple it to lower-spin fields introducing gauge invariances that may restore unitarity, and whether these putative theories provide a phenomenologically acceptable description of $\Delta$ resonances.

\section{Acknowledgments}

The author wishes to thank Holger Gies for discussing many delicate points of the present work throughout the past months. Furthermore, the author is grateful to Dario Zappalà for his hospitality at the INFN section of Catania in the last stages of the preparation of this paper. Finally, the author thanks Marco Piva for providing useful comments after the preprint version of the manuscript. 

The work of the author is supported by a Della Riccia foundation grant. The author is also thankful to the Theoretisch-Physikalisches Institut of the Friedrich-Schiller-Universität of Jena for its hospitality and financial support.

%%%%%%%%%%%%%%%%%%%%%%%%%%%%%%%%%%%%%%%%%%%%

%\appendix

%%%%%%%%%%%%%%%%%%%%%%%%%%%%%%%%%%%%%%%%%%%%

\appendix  % label section numbers alphabetically: "A", "B", etc
\counterwithin*{equation}{section} % reset 'equation' counter whenever '\section' is executed
\renewcommand\theequation{\thesection\arabic{equation}}

\renewcommand{\theequation}{A.\arabic{equation}}
\setcounter{equation}{0}

\section*{Appendix}
\setcounter{section}{0}

\subsection{Algebraic relations of gamma matrices}\label{sect:app:gamma-matrices}

In this Appendix the conventions are fixed and some useful properties of gamma matrices are collected.

The flat Minkowski metric is $\eta_{\mu\nu}=diag(+,+,+,-)$. The pseudoscalar generator of the Clifford algebra is chosen to be given by
\begin{equation}\label{eq:def-gamma5}
	\gamma_5 \equiv \frac{i}{4!} \varepsilon^{\mu\nu\rho\sigma} \gamma_\mu \gamma_\nu \gamma_\rho \gamma_\sigma \, ,
\end{equation}
yielding the following set of generators of the Clifford algebra:
\begin{equation}\label{eq:clifford-basis}
	\id \, , \quad \gamma_5 \, , \quad \gamma_\mu \, , \quad \gamma_5 \gamma_\mu \, , \quad \gamma_{\mu\nu} \equiv \gamma_{[\mu} \gamma_{\nu]} \, .
\end{equation}
We dub these generators as scalar, pseudoscalar, vector, pseudovector and tensor due to their well-known transformation properties under the Lorentz group.

The explicit representation of the Clifford algebra is given by
\begin{eqnarray}\label{eq:gamma0-&-gammai}
		\gamma^0 = -i \begin{pmatrix}
				0 & \id \\
				\id & 0 
			\end{pmatrix} \, , \quad	&&	
			\gamma_i = -i \begin{pmatrix}
				0 & \sigma_i \\
				-\sigma_i & 0 
			\end{pmatrix} \, ,
\end{eqnarray}
where $\sigma_i$ are the Pauli matrices.

We also list some useful identities, which are heavily employed in the main text. We start by considering the product of three vector generators, which yields
\begin{equation}\label{eq:gamma1-3-times}
	\gamma^\mu \gamma^\nu \gamma^\rho = i \varepsilon^{\mu\nu\rho}{}_\lambda \gamma_5 \gamma^\lambda  + g^{\mu\nu} \gamma^\rho - g^{\mu\rho} \gamma^\nu + g^{\nu\rho} \gamma^\mu \, .
\end{equation}
From this equation we readily see how to express the products of the vector and tensor generators, i.e.,
\begin{subequations}
	\begin{align}\label{eq:gamma1-gamma2}
		& \gamma^\lambda \gamma^{\alpha\beta} = g^{\lambda\alpha} \gamma^\beta - g^{\lambda\beta} \gamma^\alpha + i \varepsilon^{\lambda\alpha\beta\rho} \gamma_5 \gamma_\rho \, , \\\label{eq:gamma2-gamma1}
		& \gamma^{\alpha\beta} \gamma^\lambda = g^{\lambda\beta} \gamma^\alpha - g^{\lambda\alpha} \gamma^\beta + i \varepsilon^{\alpha\beta\lambda\rho} \gamma_5 \gamma_\rho \, .
	\end{align}
\end{subequations}
Next, we take into account the product of the pseudoscalar and tensor generators, which depends on the tensor generator only, in formula
\begin{equation}\label{eq:gammastar-gamma2}
	 \gamma_5 \gamma_{\mu\nu} = - \frac{i}{2} \varepsilon_{\mu\nu}{}^{\alpha\beta} \gamma_{\alpha\beta} \, .
\end{equation}
Another useful relation is provided by reducing the product of four vector generators, which yields
\begin{align}\label{eq:gamma1-4-times}
	\gamma_\mu \gamma_\nu \gamma_\rho \gamma_\sigma = & \left( g_{\mu\nu} g_{\rho\sigma} - g_{\mu\rho} g_{\nu\sigma} + g_{\mu\sigma} g_{\nu\rho} \right) \id \\\nonumber
	& + i \varepsilon_{\mu\nu\rho\sigma} \gamma_5 + g_{\mu\nu} \gamma_{\rho\sigma} - g_{\mu\rho} \gamma_{\nu\sigma} + g_{\mu\sigma} \gamma_{\nu\rho}\\\nonumber
	& - g_{\nu\sigma} \gamma_{\mu\rho} + g_{\nu\rho} \gamma_{\mu\sigma} + g_{\rho\sigma} \gamma_{\mu\nu} \, .
\end{align}
From this equation we immediately find the reduction of the product of two tensor generators
\begin{align}\label{eq:gamma2-gamma2}
	\gamma_{\mu\nu} \gamma_{\rho\sigma} = & \left(g_{\mu\sigma} g_{\nu\rho} - g_{\mu\rho} g_{\nu\sigma} \right) \id + i \varepsilon_{\mu\nu\rho\sigma} \gamma_5\\\nonumber
	& - g_{\mu\rho} \gamma_{\nu\sigma} + g_{\mu\sigma} \gamma_{\nu\rho} - g_{\nu\sigma} \gamma_{\mu\rho} + g_{\nu\rho} \gamma_{\mu\sigma} \, .
\end{align}

When the dimension $d$ is generic the rank-$3$ element of the Clifford basis is written as
\begin{equation}
	\gamma^{\mu\nu\rho} \equiv \gamma^{[\mu} \gamma^\nu \gamma^{\rho]} = \frac{1}{2} \left( \gamma^\mu \gamma^\nu \gamma^\rho - \gamma^\rho \gamma^\nu \gamma^\mu \right) \, .
\end{equation}
By using the properties of the Grassmann algebra we find that
\begin{equation}\label{eq:gamma3-to-gamma1}
	\gamma^{\mu\nu\rho} = \gamma^\mu \gamma^\nu \gamma^\rho + g^{\mu\rho} \gamma^\nu - g^{\mu\nu} \gamma^\rho - g^{\nu\rho} \gamma^\mu \, .
\end{equation}

\subsection{Complete form of the coefficient functions}\label{sect:app:coeff-functions}

Here, we collect the coefficient functions that were omitted from the main text. These belong to the $j=\tfrac{3}{2}$ irreducible representation with $s\neq\tfrac{3}{2}$.

We start from the coefficient functions of the annihilation operators. For $s=\pm\tfrac{1}{2}$ they are given by
\begin{equation}\label{eq:varphi-3/2-app-1}
	\begin{split}
		& \varphi^{3/2}_{\mu}(\boldsymbol{0},\tfrac{1}{2}) = \frac{1}{2\sqrt{3}} \begin{pmatrix}
			{\small\begin{pmatrix}
				0\\
				1\\
				0\\
				1\\
			\end{pmatrix}} \, , \,  {\small\begin{pmatrix}
				0\\
				i\\
				0\\
				i
			\end{pmatrix}} \, , {\small\begin{pmatrix}
				-2\\
				0\\
				-2\\
				0
			\end{pmatrix}}  \, , 0 
		\end{pmatrix}^t \, , \\
		& \varphi^{3/2}_{\mu}(\boldsymbol{0},- \tfrac{1}{2}) = \frac{1}{2\sqrt{3}} \begin{pmatrix}
			{\small\begin{pmatrix}
				- 1\\
				0\\
				-1\\
				0
			\end{pmatrix}} \, , \,  {\small\begin{pmatrix}
				i\\
				0\\
				i\\
				0
			\end{pmatrix}} \, ,  {\small\begin{pmatrix}
				0\\
				-2 \\
				0\\
				-2
			\end{pmatrix}} \, , 0 
		\end{pmatrix}^t \, .
	\end{split}
\end{equation}
On the other hand, the $s=-\tfrac{3}{2}$ component reads
\begin{equation}\label{eq:varphi-3/2-app-2}
	\varphi^{3/2}_{\mu}(\boldsymbol{0},- \tfrac{3}{2}) = \frac{1}{2} \begin{pmatrix}
		{\small\begin{pmatrix}
			0\\
			-1\\
			0\\
			-1\\
		\end{pmatrix}} \, , \,  {\small\begin{pmatrix}
			0\\
			i \\
			0\\
			i
		\end{pmatrix}} \, , 0 \, , 0 
	\end{pmatrix}^t \, .
\end{equation}

The remaining functions are those multiplying the creation operators. In this case we have
\begin{equation}
	\begin{split}\label{eq:chi-3/2-app}
		& \chi^{3/2}_\mu(\boldsymbol{0},\tfrac{3}{2})=\frac{1}{2} \left( 
		{\small\begin{pmatrix}
			0\\
			1\\
			0\\
			-1\\
		\end{pmatrix}}\, ,{\small\begin{pmatrix}
			0\\
			-i\\
			0\\
			i\\
		\end{pmatrix}},0,0 \right)^t \, , \\
		& \chi^{3/2}_\mu(\boldsymbol{0},\tfrac{1}{2})=\frac{1}{2\sqrt{3}} \left( 
		{\small\begin{pmatrix}
			-1\\
			0\\
			1\\
			0\\
		\end{pmatrix}}\, ,{\small\begin{pmatrix}
			i\\
			0\\
			-i\\
			0\\
		\end{pmatrix}},{\small\begin{pmatrix}
			0\\
			-2\\
			0\\
			2\\
		\end{pmatrix}},0 \right)^t \, ,\\
		& \chi^{3/2}_\mu(\boldsymbol{0},-\tfrac{1}{2})=\frac{1}{2\sqrt{3}} \left( 
		{\small\begin{pmatrix}
			0\\
			-1\\
			0\\
			1\\
		\end{pmatrix}}\, ,{\small\begin{pmatrix}
			0\\
			-i\\
			0\\
			i\\
		\end{pmatrix}},{\small\begin{pmatrix}
			2\\
			0\\
			-2\\
			0\\
		\end{pmatrix}},0 \right)^t \, ,\\
		& \chi^{3/2}_\mu(\boldsymbol{0},-\tfrac{3}{2})=\frac{1}{2} \left( 
		{\small\begin{pmatrix}
			1\\
			0\\
			-1\\
			0\\
		\end{pmatrix}}\, ,{\small\begin{pmatrix}
			i\\
			0\\
			-i\\
			0\\
		\end{pmatrix}},0,0 \right)^t \, .
	\end{split}
\end{equation}

\subsection{Tensors for the $1$-loop integration}\label{sect:app:1-loop}

Here, we provide the explicit forms of the generalized curvature, the endomorphism, and the perturbation tensor $\tilde{\cal M}(2)$ that were introduced in the main text. The first one is the only one that is model independent, and it takes the following form:
\begin{equation}\label{eq:generalized-curvature}
	\begin{split}
	\Omega_{\mu\nu\,\rho}{}^\lambda = & - \frac{7}{8} R^\lambda{}_{\rho\mu\nu} + \frac{3}{16} \gamma^{\alpha\beta} R_{\alpha\beta\mu\nu} \delta_\rho{}^\lambda + \frac{1}{8} R^\alpha{}_{\rho\mu\nu} \gamma_\alpha{}^\lambda\\
	& + \frac{1}{8} \gamma_\rho{}^\alpha R^\lambda{}_{\alpha\mu\nu} + \frac{i}{16} \gamma_5 \epsilon^\lambda{}_{\rho\alpha\beta} R^{\alpha\beta}{}_{\mu\nu} \, .
	\end{split}
\end{equation}
On the other hand, the endomorphism is written as
\begin{equation}\label{eq:endomorphism}
	\begin{split}
		{\cal E}_\mu{}^\nu = & - \frac{1}{2} R_\mu{}^\nu - \frac{1}{8} R \delta_\mu{}^\nu + \frac{1}{4} \gamma_\mu{}^\alpha R^\nu{}_\alpha - \frac{1}{4} R^\alpha{}_\mu \gamma_\alpha{}^\nu\\
		& + \frac{1}{8} R \gamma_\mu{}^\nu - \frac{1}{2} R^\nu{}_{\mu\alpha\beta} \gamma^{\alpha\beta} - \frac{3}{4} m^2 \delta_\mu{}^\nu + \frac{1}{4} m^2 \gamma_\mu{}^\nu \, .
	\end{split}
\end{equation}
Finally, the tensor coefficient of the differential operator $\tilde{\cal M}(2)$ is given by
\begin{equation}\label{eq:M2-tilde}
	\begin{split}
	\tilde{\cal M}(2)^{\alpha\beta}{}_\mu{}^\nu = &  - \zeta \delta^\alpha{}_\mu R^{\beta\nu} + \frac{4 \zeta}{4 - 3 \zeta} g^{\beta\nu} R^\alpha{}_\mu\\
	& - \frac{\zeta^2}{4(4-3\zeta)} \delta^\alpha{}_\mu g^{\beta\nu} R + \frac{2\zeta}{4-3\zeta} \gamma^{\rho\sigma} g^{\beta\nu} R^\alpha{}_{\mu\rho\sigma}\\
	& + \frac{\zeta}{2} \gamma^{\rho\sigma} \delta^\alpha{}_\mu R^{\beta\nu}{}_{\rho\sigma} \, .
	\end{split}
\end{equation}

%%%%%%%%%%%%%%%%%%%%%%%%%%%%%%%%%%%%%%

\bibliographystyle{chetref}
%\bibliography{biblio}

%%%%%%%%%%%%%%%%%%%%%%%%%%%%%%%%%%%%%%%%%%%%%%%%%%%%%%%%%%

\end{document}